\documentclass[seceq]{ptptex}
\usepackage{epsfig}

\def\lsim{\raise0.3ex\hbox{$<$\kern-0.75em\raise-1.1ex\hbox{$\sim$}}}
\def\gsim{\raise0.3ex\hbox{$>$\kern-0.75em\raise-1.1ex\hbox{$\sim$}}}
\title{%        %You can use \\ for explicit line-break
Lattice QCD at non-zero chemical potential and\\ 
the resonance gas model
\vspace*{-1.7cm}
\hfill\mbox{BI-TP 2004/04}
\vspace*{1.5cm}
}

\author{%       %Use \scshape  for the family name
Frithjof \textsc{Karsch}
}

\inst{%         %Affiliation, neglected when [addenda] or [errata]
Fakult\"at f\"ur Physik, 
Universit\"at Bielefeld, 
D-33615 Bielefeld,
Germany
}

\abst{%         %this abstract is neglected when [addenda] or [errata]
We present results from lattice calculations on the thermodynamics of
QCD at non-zero temperature and baryon chemical potential and 
discuss the role of resonances for the occurrence of the transition
to the quark-gluon plasma in hot and dense matter. Properties of
a hadronic resonance gas are compared to lattice results on the
equation of state at zero as well as non-zero baryon chemical potential.
Furthermore, it is shown that the quark mass dependence of the transition
temperature can be understood in terms of lines of constant energy density
in a resonance gas.
}

\begin{document}

\maketitle

\section{Introduction}
Today properties of strongly interacting matter at high temperature and
non-zero baryon number density are analyzed within the field theoretic 
framework given by the
theory of strong interactions -- Quantum Chromo Dynamics (QCD). Two
aspects of this theory, which describes the interaction among their
elementary constituents, quarks and gluons, are of central importance for
our understanding of the different phases of QCD at non-zero temperature
and density as well as for our understanding of the experimentally
observed spectrum of hadronic bound states -- confinement and chiral
symmetry breaking. The $SU_R(n_f)\times SU_L(n_f)$ chiral symmetry,
which is spontaneously broken in QCD with $n_f$ massless quark flavors, as
well as the $Z(N_c)$
center symmetry, realized in the gauge field sector of $N_c$-color QCD
in the limit of infinitely heavy quarks, put strong constraints on
qualitative aspects of the QCD phase diagram\cite{Svetitsky,Wilczek}.
In fact, quite general considerations concerning universal properties
of the QCD phase transition rely on these symmetries. Many of the emerging
predictions concerning the order of the QCD transition and concerning
universal properties in the vicinity of second order phase transitions
in QCD have been verified over the last 20 years in numerical studies
of lattice regularized QCD\cite{lattice_review}.

%%%%%%%%%%%%%%%%%%%%%%%%%%%%%%%%%%%%%%%%%%%%%%%%%%%%%%%%%%%%%%%%%%%%%%%
\begin{figure}[t]
\epsfig{file=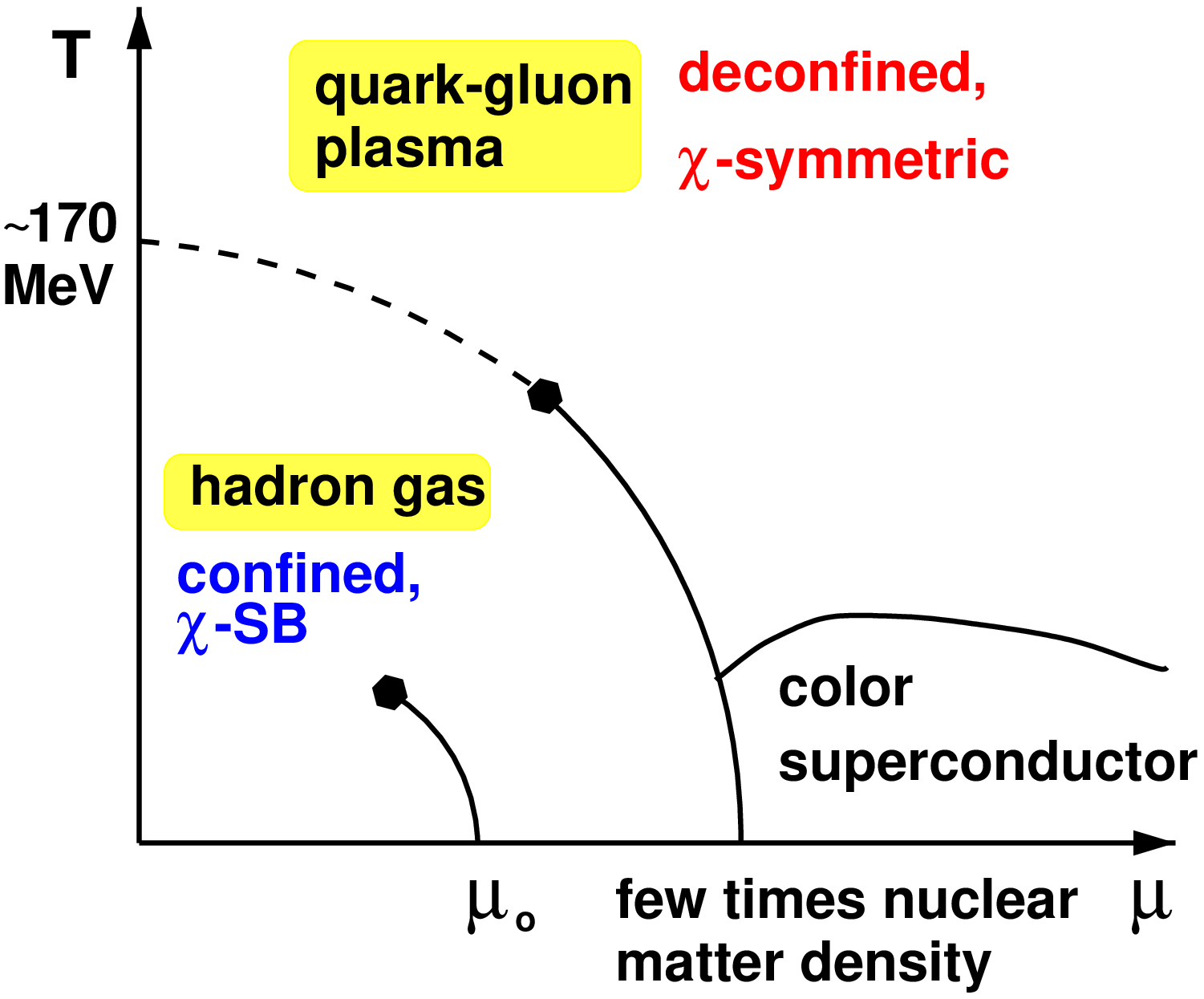,width=7.0cm}\hfill\epsfig{file=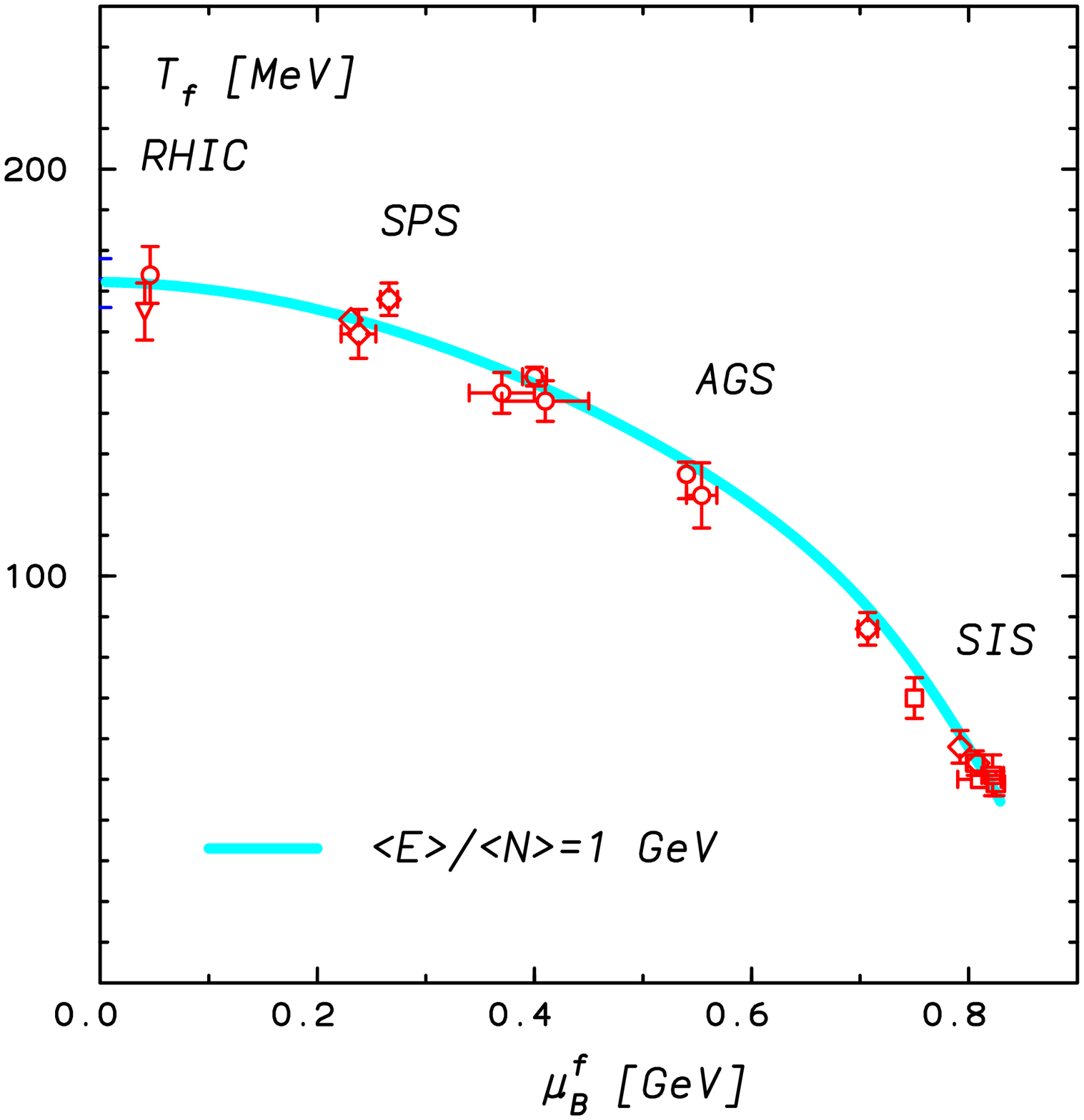,
width=6.1cm}
\caption{Sketch of the QCD phase diagram in the $T$-$\mu_B$ plane (left)
and the freeze-out
curve (right) determined from a comparison of experimentally observed
particle abundances to the abundances of hadrons in a hadronic resonance
gas at temperature $T$ and a baryon chemical potential $\mu_B$.
\label{phase}}
\end{figure}
%%%%%%%%%%%%%%%%%%%%%%%%%%%%%%%%%%%%%%%%%%%%%%%%%%%%%%%%%%%%%%%%%%%%%%%

In our attempt to reach a deeper understanding of the physics behind
the occurrence of the QCD phase transition, mechanisms like the dual
Higgs mechanism, monopole 
condensation or vortex percolation have been
identified\cite{understanding}, 
which characterize the drastic modifications of the QCD 
vacuum that occur at the critical temperature. However, none of these
considerations provides insight into the question which properties of QCD 
set the scale and control quantitative aspects of 
the transition from hadronic matter at low
temperature and density to the quark gluon plasma (QGP) at high temperature
and/or density. 
%Resonance gas models may give insight into this question. 

Long before lattice calculations provided first evidence for a phase transition
in strongly interacting matter\cite{first} the inevitable need for critical 
behavior in hadronic matter has been discussed in the 
framework of resonance gas models\cite{Hagedorn}. It has been 
noticed that ordinary hadronic matter cannot persist  
at arbitrary high temperatures and densities\cite{Hagedorn}; the copious
production of resonances will lead to a natural end of the temperature
and density regime in which hadrons can exist. It also   
has been suggested that the properties of dense hadronic matter and the 
role of resonances could be studied experimentally
in heavy ion collisions\cite{Teller}. Indeed, in such experiments it has 
been found  that the abundances of various particle species  
are well described by a hadronic resonance gas model\cite{Redlich} 
which is characterized by two equilibrium parameters, temperature, $T$,
and baryon chemical potential, $\mu_B$. Both depend on the
center of mass energy in these collisions; with increasing energy the 
temperature of the resonance gas increases and the relevant baryon 
chemical potential decreases. At RHIC the baryon chemical potential is
quite small ($\mu_B \simeq 29$~MeV) while the temperature of the resonance 
gas reaches 177~MeV\cite{Braun-Munzinger}. This is in good agreement
with the transition temperature to the quark-gluon plasma phase, 
$T_c = (173 \pm 8 \pm sys)$~MeV, found in lattice calculations\cite{Peikert}
of 2-flavor QCD at vanishing baryon chemical potential $\mu_B$. 
Recent exploratory
lattice calculations also show that the transition temperature drops slowly 
with increasing baryon chemical potential\cite{Fodor1,Allton1,deForcrand}
and that the smooth transition at small $\mu_B$ turns into a second
order transition at a critical point\cite{Fodor1}  
$(T_c,\mu_B^c)$ where $T_c$ is about 10\% smaller than the transition
temperature at $\mu_B=0$ and estimates for $\mu_B^c$ range from
$\mu_B^c \simeq  725~$~MeV\cite{Fodor1} to $\mu_B^c \simeq 420$~MeV\cite{Ejiri}. 
The anticipated QCD phase diagram
and the freeze-out curve determined from the particle abundances 
observed in various heavy ion experiments are shown in Fiq.~\ref{phase}.

In this paper we want to discuss evidence provided by lattice
calculations for the contribution of hadron resonances to the 
thermodynamics of QCD. We will analyze the equation of state and
the quark mass dependence of the QCD transition temperature. 
Furthermore, we will address the question to what extend the freeze out
temperature observed in heavy ion experiments is related to the 
phase boundary for the transition to the QGP calculated in lattice QCD.

\section{Lattice results on $T_c$ and the equation of state}

\subsection{$\mu_B \; =\; 0$}

%%%%%%%%%%%%%%%%%%%%%%%%%%%%%%%%%%%%%%%%%%%%%%%%%%%%%%%%%%%%%%%%%%%%%%%
\begin{figure}[t]
\epsfig{file=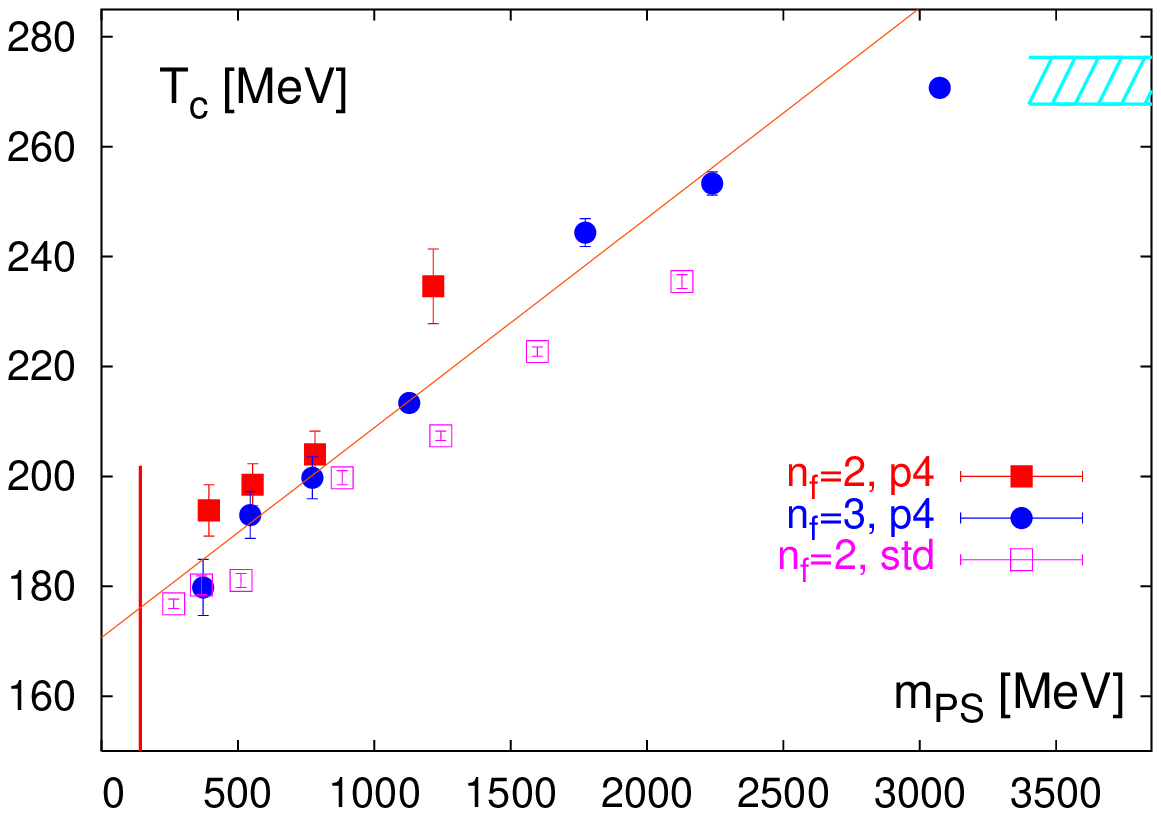,width=6.6cm}
\hfill 
\epsfig{file=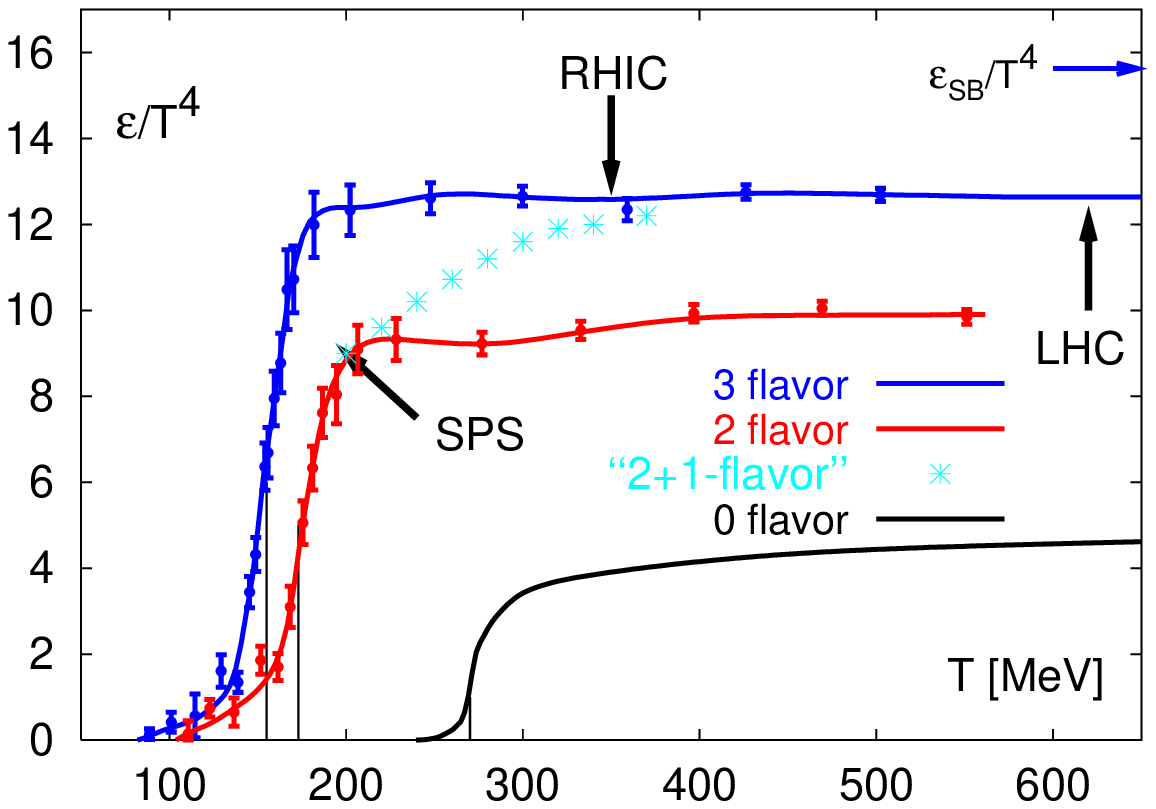,width=6.6cm}
\caption{The transition temperature (left) as a function of the lightest
pseudo-scalar meson mass ($m_{PS}$) in 2 and 3 flavor QCD and the energy 
density (right) in the SU(3) gauge theory ($n_f=0$), as well as 2 and 3 
flavor QCD for quark masses corresponding to $m_{PS} \simeq 770$~MeV.
\label{qcd}} 
\end{figure}
%%%%%%%%%%%%%%%%%%%%%%%%%%%%%%%%%%%%%%%%%%%%%%%%%%%%%%%%%%%%%%%%%%%%%%%

Studies of the quark mass dependence of the transition to the high
temperature phase of QCD show that the transition temperature
decreases gradually with decreasing quark mass\footnote{Quark masses
used in lattice calculations are bare parameters of the QCD Lagrangian
which need to
be renormalized. In order to avoid any discussion of the renormalization 
of quark masses it is more appropriate to discuss the quark mass dependence
of thermodynamic observables in terms of a well controlled physical parameter.
E.g. we will use here the lightest pseudo-scalar meson (pion) mass, $m_{PS}$. 
To express this in physical units (MeV) we use zero temperature lattice  
calculations of the string tension and set the scale by using $\sqrt{\sigma}
= 420$~MeV.}. Over a wide range of pion mass values, $300~{\rm MeV}\; \lsim\;
m_{PS}\; \lsim\; 2~{\rm GeV}$, the transition temperature
depends linearly on $m_{PS}$ and the slope seems to be more or less
independent of the number of flavors (Fig.~\ref{qcd}(left)). 
In Ref.~10 the mass dependence of $T_c$ has been parametrized as,
\begin{equation}
T_c(m_{PS}) = T_c (0) + 0.04 (1) m_{PS} \quad .
\label{Tcm}
\end{equation}
For smaller pion masses one may expect that chiral symmetry leads to 
modifications of this linear relation\footnote{In the case of a second
order transition in 2-flavor QCD one expects to find for small values of
the pseudo-scalar mass, $T_c \; \sim\; m_{PS}^{2/\beta\delta}$, where 
$2/\beta\delta\; \simeq\; 1.1$
is a combination of critical exponents of the 3-$d$, $O(4)$-model.}; for 
$m_{PS} \; \gsim\; 2~$GeV the transition temperature will approach a constant 
as all meson and baryon masses become larger than the (almost) quark mass
independent glueball-masses and thus will decouple from the thermodynamics. 

The change in transition temperature between the light quark mass and 
infinite quark mass regime goes along with a change in $\epsilon_c/T_c^4$,
the critical energy density expressed in units of the transition
temperature, by more than an order of magnitude (Fig.~\ref{qcd}(right)), 
{\it i.e.} $\epsilon_c/T_c^4 \simeq (6\pm 2)$ for $m_{PS} \simeq 770$~MeV 
and $\epsilon_c/T_c^4 \simeq (0.5-1)$ for $m_{PS} \equiv \infty$. This large
change reflects the large difference in the number of degrees of freedom
which control the high temperature ideal gas limit. 
However, it does not at all suggest that the critical energy density itself 
changes significantly.  In fact, when
taking into account the shift in $T_c$ for both cases it seems that the
critical energy density itself does not change much. Although statistical
errors are still large both cases are consistent with a critical
energy density in the range $\epsilon_c \simeq (0.5 - 1.0)$~GeV/fm$^3$.

\subsection{$\mu_B\; >\; 0$}

Lattice simulations at non-zero baryon chemical potential\footnote{We use
here the notion of baryon chemical potential $\mu_B$ although lattice
calculations are performed in terms of the quark chemical potential 
$\mu_q\equiv \mu_B/3$.} generally
suffer from the problem that the fermion contribution to the QCD
partition function is no longer represented by a positive definite
quantity; the fermion determinant becomes complex. This excludes the
application of standard Monte Carlo simulation techniques\cite{Nakamura}. 
This problem has, however, been avoided in recent studies of the dependence 
of the transition
temperature on the chemical potential\cite{Fodor1,Allton1,deForcrand} as
well as calculations of the equation of state\cite{Fodor2,Allton2} at 
non-zero baryon chemical potential by using extrapolation techniques 
applied to numerical results obtained at $\mu_B=0$.  

%%%%%%%%%%%%%%%%%%%%%%%%%%%%%%%%%%%%%%%%%%%%%%%%%%%%%%%%%%%%%%%%%%%%%%%
\begin{figure}[t]
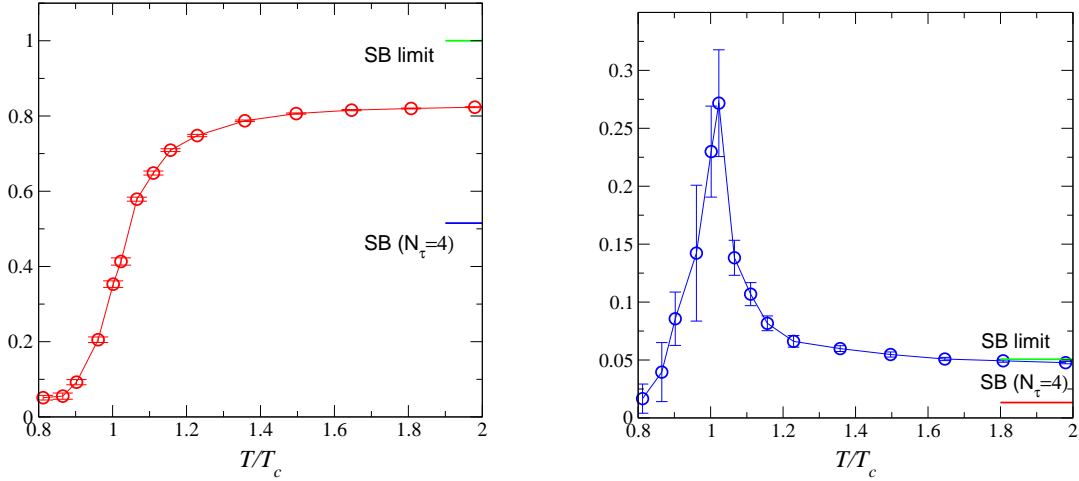

\hspace*{-0.2cm}\epsfig{file=p2.eps,width=64mm}
\hfill\epsfig{file=p4.eps, width=64mm}
\caption{The expansion coefficients $c_2$ and $c_4$ calculated in 
2-flavor QCD for several values of the temperature. 
As in the $\mu_B=0$ case the simulations have 
been performed with quark masses corresponding to $m_{PS} \simeq 770$~MeV
on a lattice of size $16^3\times 4$.
\label{c2c4}}
\end{figure}
%%%%%%%%%%%%%%%%%%%%%%%%%%%%%%%%%%%%%%%%%%%%%%%%%%%%%%%%%%%%%%%%%%%%%%%

\begin{figure}[t]
\hspace*{-0.2cm}\epsfig{file=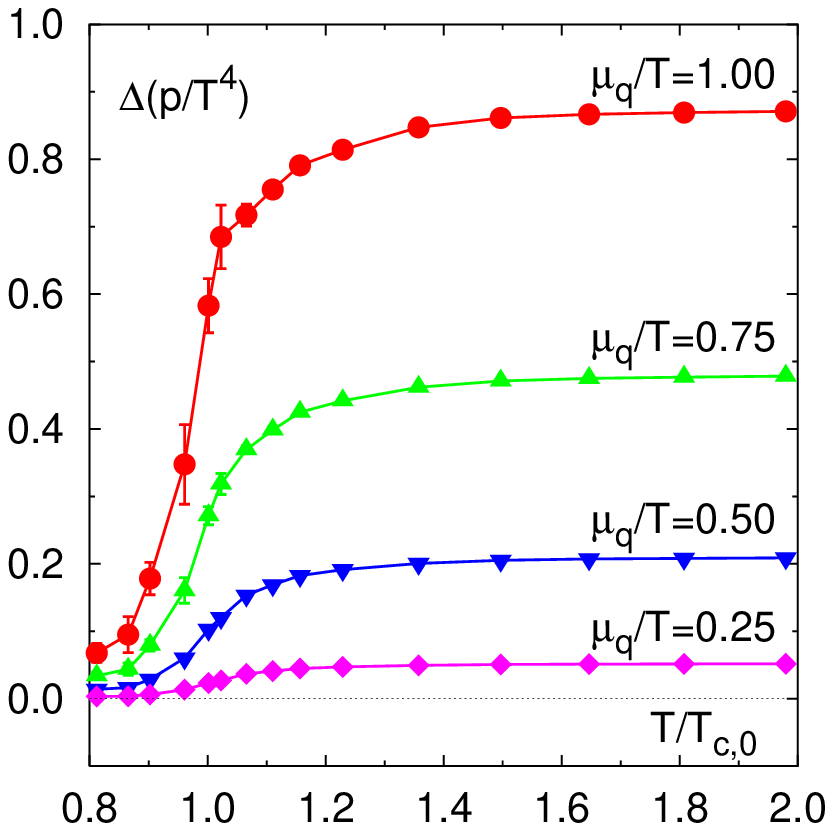,width=64mm}
\hfill\epsfig{file=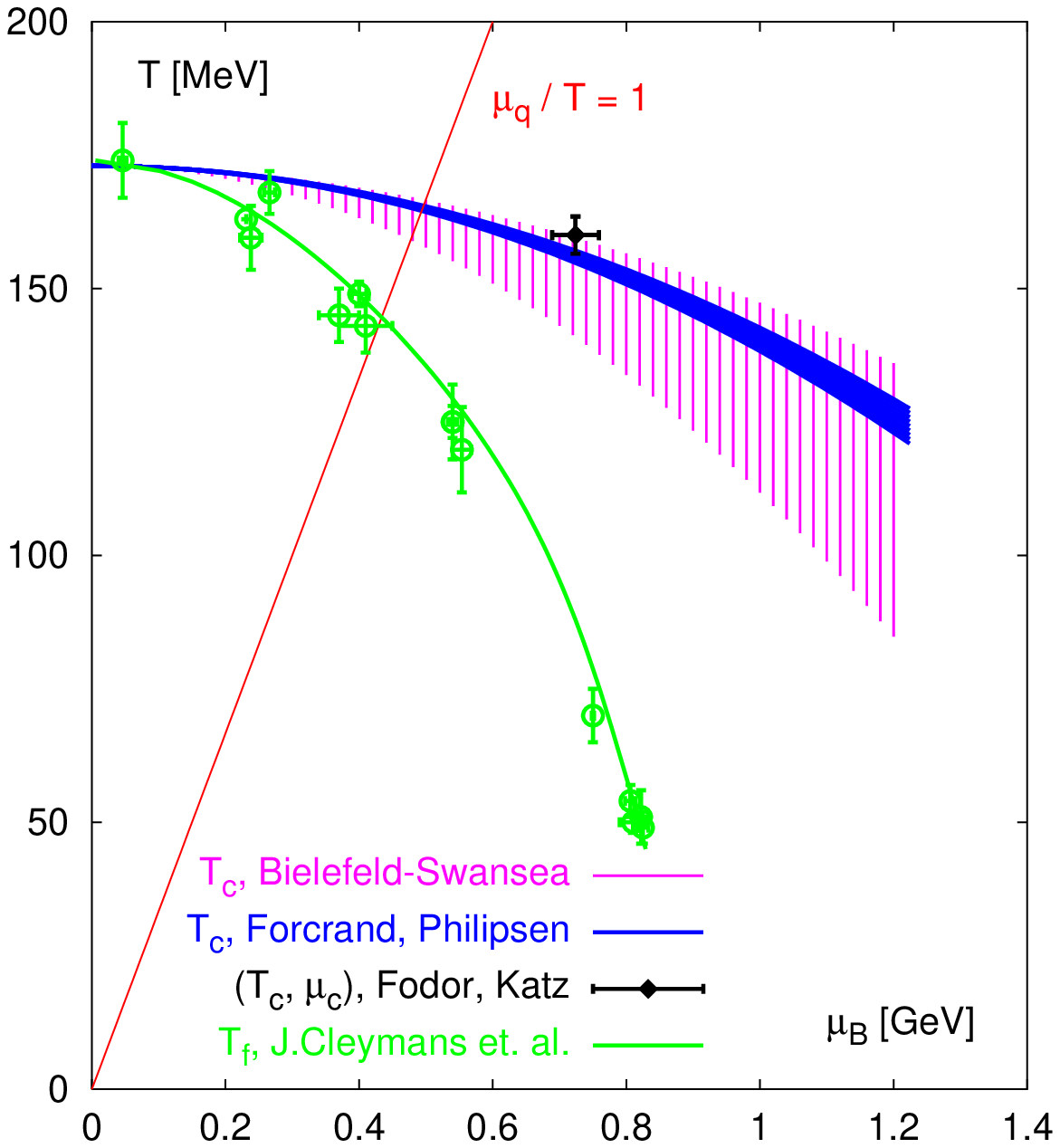, width=62mm}
\vspace*{0.2cm}
\caption{The left hand part of the figure shows the change in pressure due 
to a non-zero quark chemical potential,
$\mu_q = \mu_B/3$ calculated in 2-flavor QCD using a Taylor expansion
to order $(\mu_q/T)^4$. As in the $\mu_B=0$ case the simulations have 
been performed with quark masses corresponding to $m_{PS} \simeq 770$~MeV.
The right hand part gives results on the $\mu_B$-dependence of the transition
line together with the chemical freeze-out line of Fig.~1 which corresponds to 
a line of constant energy per particle in a 
a hadronic resonance gas.
\label{fdqcd}}
\end{figure}

At fixed temperature and small values of the chemical potential the pressure 
may be expanded in a Taylor series around $\mu_q = 0$,
\begin{equation}
{p\over T^4}={1\over{VT^3}}\ln{\cal Z} = \sum_{n=0}^{\infty} c_n(T)\quad ,
\label{Taylorp}
\end{equation}
where the expansion coefficients are given in terms of derivatives of 
the logarithm of the QCD partition function, $c_n(T) = \displaystyle{\frac{1}{n!}
\frac{\partial^n \ln Z}{\partial (\mu_q / T)^n}}$.  The series is an
even series in $(\mu_q/T)$. The first coefficient, $c_0$,
just gives the pressure studied for some time in finite temperature
lattice calculations at $\mu_q = 0$. The first non-zero coefficient, $c_2$,
is proportional to the quark number susceptibility at $\mu_q = 0$ which 
also has been studied in the past\cite{Gottlieb}. The relevant expansion
coefficients entering the calculation of the pressure and quark number 
susceptibility in a Taylor expansion of the partition function up to 
${\cal O} (\mu_q^4)$\cite{Allton2},
\begin{eqnarray}
\frac{\Delta p(T,\mu_q)}{T^4} &=& \frac{p(T,\mu_q) - p(T,0)}{T^4} =
c_2 \biggl( {\mu_q \over T} \biggr)^2
+c_4 \biggl( {\mu_q \over T} \biggr)^4 \nonumber \\
{\chi_q \over T^2} &=& 2 c_2 + 12 c_4 \biggl( {\mu_q \over T}\biggr)^2 \quad ,
\label{taylor2}
\end{eqnarray}
are shown in Fig.~\ref{c2c4}.
In Fig.~\ref{fdqcd}(left) we show corresponding results on the change in 
pressure, Similar results have been obtained 
using a reweighting technique\cite{Fodor2}. A comparison with the pressure 
calculated at $\mu_q = 0$ shows that at $\mu_q/T =1$ and for $T\; \gsim\; T_c$
the enforced presence of a non-zero baryon number adds about 30\% to the 
overall pressure in the system. This also confirms that the Taylor expansion
converges rapidly up to $\mu_q/T \sim 1$. 

%%%%%%%%%%%%%%%%%%%%%%%%%%%%%%%%%%%%%%%%%%%%%%%%%%%%%%%%%%%%%%%%%%%%%%%
%%%%%%%%%%%%%%%%%%%%%%%%%%%%%%%%%%%%%%%%%%%%%%%%%%%%%%%%%%%%%%%%%%%%%%%

The reweighting 
approach\cite{Fodor1,Allton1} as well as the analytic continuation of results 
obtained in simulations with an imaginary chemical potential\cite{deForcrand}
have also been used to determine the $\mu_B$-dependence of the transition
temperature. These results are shown in Fig.~\ref{fdqcd}(right). We note that
for small values of the chemical potential the lines shown in this 
figure do not correspond to a 
phase transition but rather characterize a rapid but smooth crossover from
the hadronic to the plasma phase. Also shown in the figure is an estimate
of the chiral critical point\cite{Fodor1}, {\it i.e.} a second order phase 
transition point, 
$(T_c, \mu_B^c)\simeq (160~{\rm MeV}, 725~{\rm MeV})$, 
in the QCD phase diagram. For $\mu_B > \mu_B^c$ the transition is expected
to become a first order phase transition.   

Although the different methods used to determine $T_c(\mu_B)$ do seem to
give compatible results, a detailed quantitative comparison is difficult
as all calculations performed so far have used different discretization 
schemes and/or quark mass values. 
Moreover, the estimates have partly been obtained on rather small
lattices with unimproved gauge and fermion actions and/or too large
quark masses. The current quantitative results thus need to be improved
and confirmed in future calculations. A first analysis of the quark mass 
dependence of the transition line for 3-flavor QCD\cite{lat03} seems to 
indicate that the $\mu_B$-dependence becomes stronger with decreasing quark 
mass and that the chiral critical point in the physically realized case of 
(2+1)-flavor QCD shifts to smaller values of the baryon chemical 
potential\cite{Ejiri}, $\mu_B^c \simeq 420$~MeV.

\section{Thermodynamics of the hadronic resonance gas}

\subsection{$\mu_B\; =\; 0$}

In Hagedorn's approach to the thermodynamics of strongly interacting
matter\cite{Hagedorn} critical behavior arises because any increase
in energy of an ensemble of strongly interacting hadrons is
predominantly used to generate new resonances rather
than transforming it into kinetic energy of the constituents. Energy thus
is not used for heating up the system any further. The exponentially rising 
spectrum of resonances, $\rho (m) \sim \exp{(b m)}$, leads
to the occurrence of a critical (limiting) temperature, $T_{\rm c,res} = 1/b$.

A similar mechanism is known to lead to critical behavior in purely
gluonic systems, {\it i.e.} in the $SU(N_c)$ gauge theories. Here the
fluctuations of color flux tubes (string) lead to an exponentially rising
excitation spectrum which again leads to critical behavior. Calculations
of the critical temperature, $T_{\rm c, string}$, within the Nambu-Goto
model yield\cite{Alvarez},
\begin{equation}
\frac{T_{\rm c, string}}{\sqrt{\sigma}} \; =\; \sqrt{\frac{3}{(d-2)\pi}}\quad ,
\label{string}
\end{equation} 
which only depends on the space-time dimension, $d$, and, in particular,
is independent of the color degrees of freedom. Lattice calculations
of the phase transition temperature of $SU(N_c)$ gauge theories in
3 and 4 space-time dimensions, indeed, yield critical temperatures,
which are in good agreement with the string model predictions. Some
results from lattice calculations\cite{TcSUN} are summarized in 
Table~\ref{tab:string} and compared with the string model prediction.
\begin{table}[t]
\caption{Lattice results on the deconfinement temperature in units of the
square root of the string tension of $SU(N_c)$
gauge theories in $d$ space-time dimensions. The last row gives results
obtained from a large-$d$ analysis of string models\cite{Alvarez}.}
\vspace*{0.2cm}
\begin{center}
{
%\footnotesize
\begin{tabular}{@{}cccl||cclc@{}}
\hline
{} &{} &{} &{} &{}&{} &{} &{} \\[-1.5ex]
{} & $d$ & $N_c$ & $T_{c}/\sqrt{\sigma}$& $d$ & $N_c$ &
$T_{c}/\sqrt{\sigma}$ &{} \\[1ex]
\hline
{} &{} &{} &{} &{} &{} &{} &{} \\[-1.5ex]
{} &3 & 2 &1.08(1) & 4  & 2 &0.69(2) &{} \\[1ex]
{} &{} & 3&0.97(1) &{} & 3 &0.632(2)  &{}  \\[1ex]
\hline
{} &{} &{} &{} &{} &{} &{} &{} \\[-1.5ex]
{} & string&model:&0.977 &string&model:&0.691 &{} \\[1ex]
\hline
\end{tabular}\label{tab:string} }
\end{center}
%\vspace*{-13pt}
\end{table}
This clearly suggests that resonances play an essential role in
triggering the occurrence of the deconfinement transition, although
the order of the transition and, in those cases where the transition is
second order, also universal properties of the transition are 
controlled by the global $Z(N_c)$ center symmetry of the $SU(N_c)$ 
gauge theories. It thus is interesting to explore what role the (exponentially)
rising hadronic resonance spectrum plays for the occurrence of the
transition to the plasma phase in the physical, light quark mass regime.

A first hint at the importance of resonances for the occurrence of the QCD 
transition may be obtained from the energy density at $T_c$. In the
limit of $n_f$ massless quark flavors the sector of massless Goldstone 
bosons would contribute to the energy density with 
$\epsilon /T_c^4 = (n_f^2-1)\pi^2/30$, if this contribution can
be approximated by an ideal gas of non-interacting bosons. The calculations 
of the energy 
density shown in Fig.~\ref{qcd}(right) have been performed with quark masses
which correspond to a pion mass of about $770$~MeV. Their contribution
to the energy density thus is exponentially suppressed. A free gas of
massive relativistic particles would contribute 
\begin{equation}
{{\epsilon^1 (m)}\over {T^4}} = \frac{g}{2\pi^2}\;
\sum_{k=1}^{\infty} \;(-\eta)^{k+1}\;\frac{1}{k}\left( \frac{m}{T} \right)^3
\;\left[\frac{3 T\;K_2(k m/T)}{k m} + \;K_1(k m/T)\right]
\label{gas}
\end{equation} 
to the energy density. Here $\eta= -1$ for bosons, $+1$ for fermions
and $g$ is the degeneracy factor of the particle state. For $m/T_c\simeq 4$ 
this yields $\epsilon/T_c^4 \simeq 0.083 (n_f^2-1)$ which shows that quite
a few hadronic degrees of freedom are needed to saturate the value of
the energy density found in lattice calculations at $T_c$.

\begin{figure}[t]
\epsfig{file=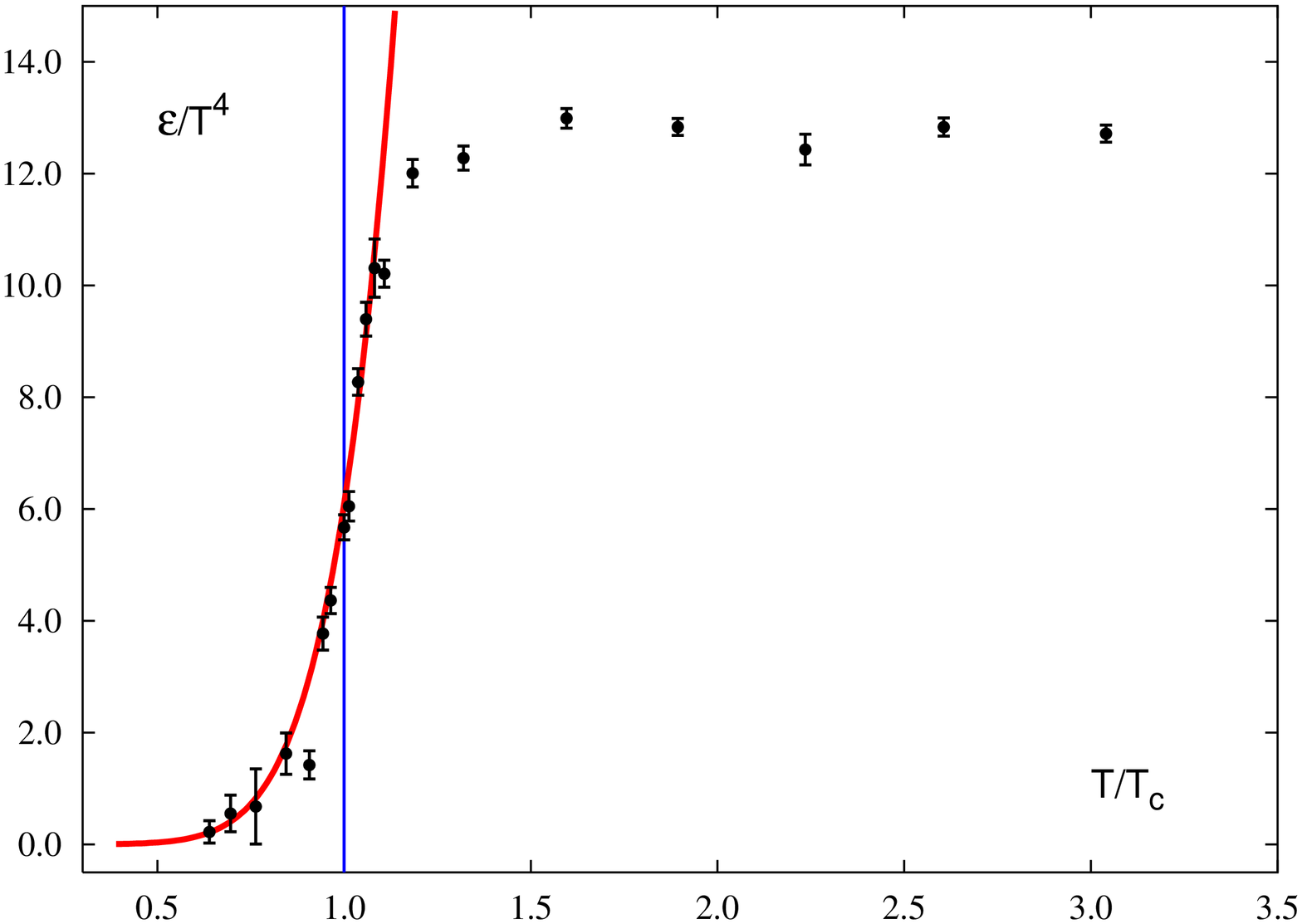,width=4.9cm,angle=-90}
\hfill
\epsfig{file=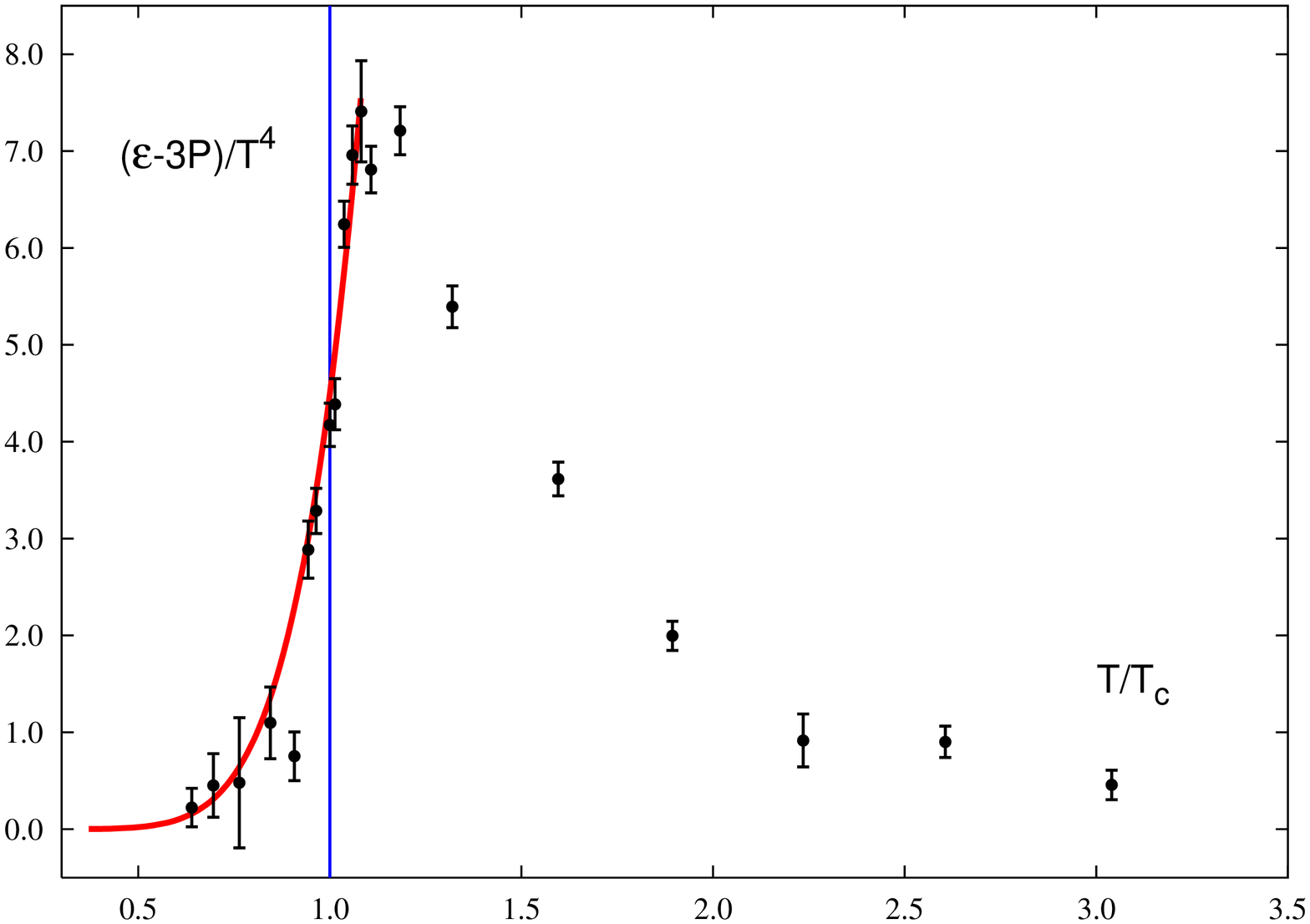,width=4.9cm,angle=-90}
\vspace*{0.2cm}
\caption{The energy density in (2+1)-flavor QCD (left) and the 
deviation from ideal gas behavior, $(\epsilon -3p)/T^4$, (right)
as a function of $T/T_c$. The solid line gives the 
result obtained from a resonance gas with hadron masses adjusted
to compare with the lattice calculations which have been performed with
too heavy up and down quarks\cite{res1}.
\label{lat_res}}
\end{figure}

In a gas of non-interacting resonances energy density and pressure
are given as sum over the single particle contributions, e.g. 
$\epsilon = \sum_{i} \epsilon^1 (m_i)$, where $i$ labels the masses, $m_i$, 
of experimentally known mesons and baryons. In order to compare 
the resonance gas model with lattice results we, however, have to take 
into account that the latter are not obtained from calculations
performed with the physically realized
light quark mass spectrum. A corresponding analysis has been performed
in Refs. 22,23. The resulting comparison between a modified resonance gas
model and lattice data for (2+1)-flavor QCD\cite{Peikert} is shown in 
Fig.~\ref{lat_res}. Similar agreement has been obtained for 2-flavor
QCD by suppressing the contribution of ''strange'' hadrons. This shows
that resonance can account for the rapid rise of the energy density observed
in lattice calculations for $T\lsim T_c$.

%%%%%%%%%%%%%%%%%%%%%%%%%%%%%%%%%%%%%%%%%%%%%%%%%%%%%%%%%%%%%%%%%%%%%%%
%%%%%%%%%%%%%%%%%%%%%%%%%%%%%%%%%%%%%%%%%%%%%%%%%%%%%%%%%%%%%%%%%%%%%%%

\subsection{$\mu_B\; >\; 0$}

\begin{figure}[t]
\epsfig{file=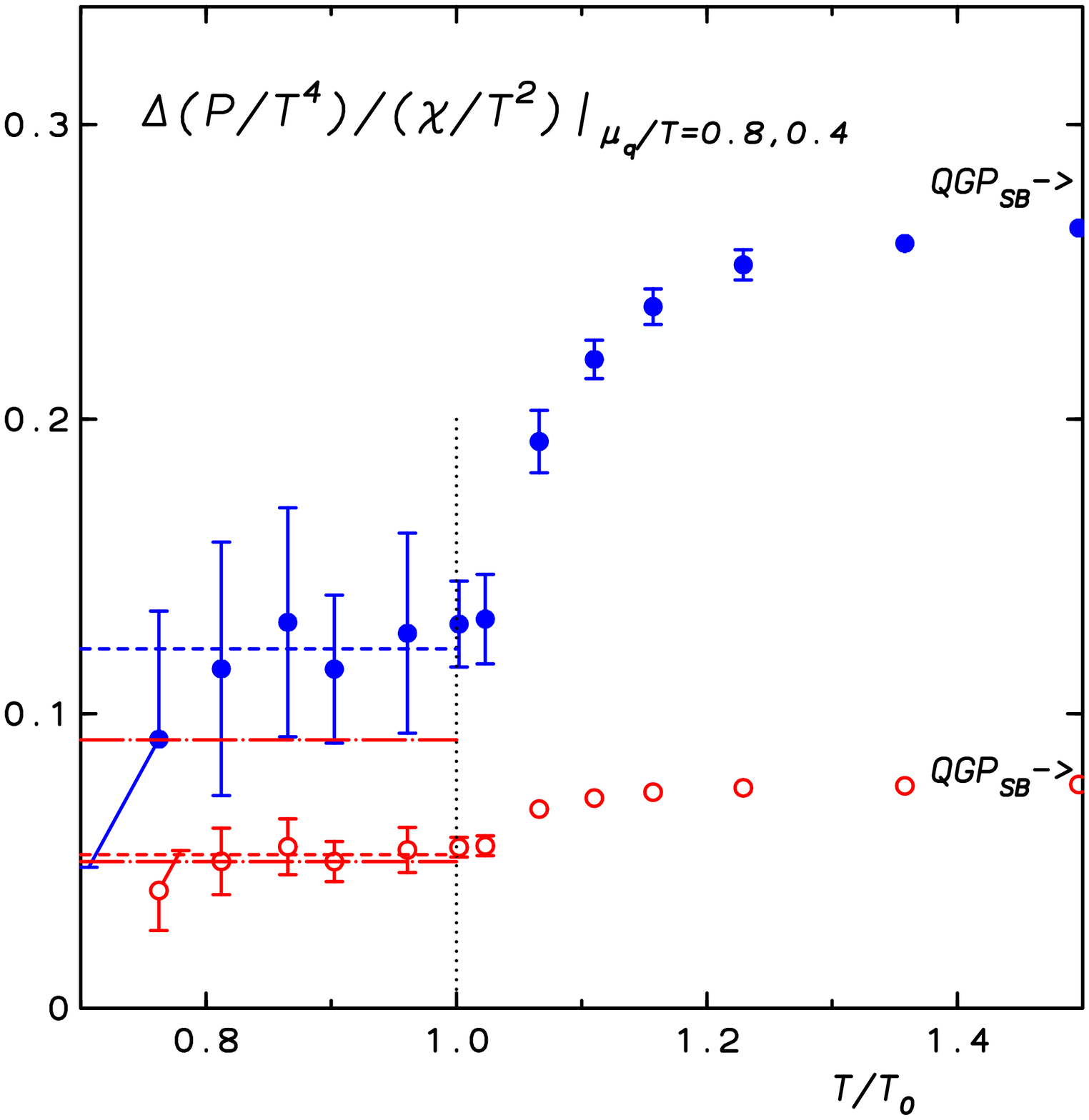,width=64mm,angle=180}
\hfill\epsfig{file=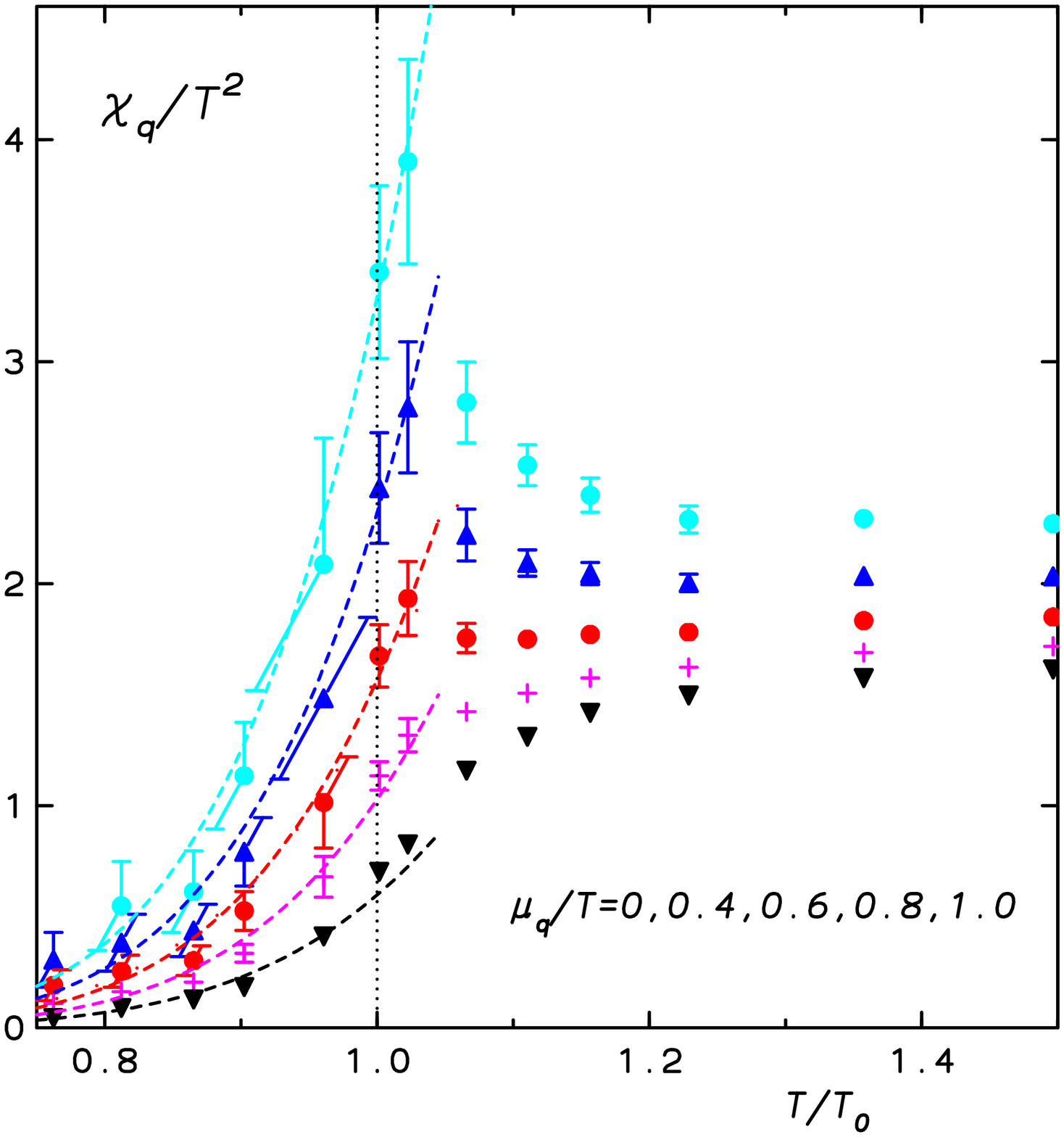,width=62mm,angle=180}
\vspace*{-0.5cm}
\caption{The ratio of pressure
and quark number susceptibility $\chi_q \equiv 9\chi_B$ 
versus temperature for fixed
values of the quark chemical potential $\mu_q / T = \mu_B / 3T$ (left). 
The horizontal lines are the results of hadron resonance gas model 
calculations\cite{res2}. The points are the lattice data from Ref.~17. While the 
dashed-dotted curves
represent the complete expression given in Eq.~\protect{\ref{sus}}
the dashed curves give the result of a Taylor expansion performed to the
same order as that used in the lattice calculations.
The right hand part of the figure shows the quark number susceptibility 
versus $T/T_c$ for different values of the quark chemical potential.
The lines give results from the resonance model calculation expanded
in a Taylor series and truncated at the same order as used in the lattice 
calculation. 
\label{p_chi}}
\end{figure}

The resonance gas model does make quite stringent predictions for
the thermodynamics at non-vanishing chemical potential\cite{res1,res2}. As all
baryons in QCD are heavy compared with the temperature regime of
interest, {\it i.e.} $m_{\rm baryon}/T \; \gsim\; 5$ for $T \le T_c$,
the contribution of baryons to the thermodynamics can be handled 
in the Boltzmann approximation. This leads to a factorization of
the temperature and fugacity, $\exp ( \mu_B/T)$, 
dependence of thermodynamic observables.
The pressure of a gas of baryons and their resonances thus can be 
written as 
\begin{equation}
{p_B(T,\mu_B) \over T^4} = F_B(T) \, \cosh (\mu_B/T)\quad ,
\label{totalp}
\end{equation}
where $F_B(T)$ is given by a sum over all baryons and their resonances,
\begin{equation}
F_B(T)\equiv \sum_i {g_i\over {\pi^2}}\left( {m_i\over T} \right)^2
K_2( {m_i/ T} )
\quad .
\label{FT}
\end{equation} 
As the mesonic part of a gas of non-interacting resonances does not
depend on the baryon chemical potential the change in pressure due 
to a non-vanishing chemical potential is entirely determined by the
baryonic sector, $\Delta p = p(T,\mu_B) - p(T,0) \equiv p_B(T,\mu_B)-
p_B(T,0)$. 
This, of course, also holds for derivatives with respect to $\mu_B$ at
fixed temperature. The factorization of the $T$ and $(\mu_B/T)$-dependent
terms then leads to simple relations among various thermodynamic
observables. For instance, one finds for the baryonic susceptibility,
\begin{eqnarray}
{\chi_B  \over T^2} &=& \left( \frac{\partial^2}{\partial (\mu_B/T)^2}
\frac{p(T,\mu_B)}{T^4} \right)_{T~fixed} \nonumber \\
&=& \frac{\Delta p}{T^4} \left( 1- \cosh^{-1}(\mu_B/T) \right)^{-1}
\quad . 
\label{sus}
\end{eqnarray}
This relation is particularly interesting as it suggests that the
ratio $\Delta p/ T^2\chi_q\equiv \Delta p/ 9T^2\chi_B$ does not depend on 
details of the hadron mass spectrum and thus can directly be compared
with lattice calculations performed with un-physically large quark masses.
Such a comparison is shown in
Fig.~\ref{p_chi}(left). The agreement between the lattice results and 
the resonance gas model relation is quite striking, although it has
to be noted that the current lattice analysis has been performed using a
Taylor series expansion to order $\mu_B^4$ only. This truncation, of course,
eliminates any true density driven singular behaviour and, for instance,
cannot lead to a divergent susceptibility. Large fluctuations which could occur 
in the vicinity of a phase transition are suppressed. This also is true for the
resonance gas model approach, which does not lead to true singular behaviour
as long as it is formulated in terms of a finite set of resonances. 

The resonance gas model also yields a reasonably good description
of the temperature dependence of thermodynamic quantities at fixed
fugacity.  This, of course, involves information on the explicit form
of the resonance spectrum. Taking again into account that lattice 
simulations so far have been performed with unphysical quark mass
values one can directly compare the quark number susceptibilities
calculated in the resonance gas model and on the lattice. This is
shown in Fig.~\ref{p_chi}(right).

\section{The critical temperature}

%%%%%%%%%%%%%%%%%%%%%%%%%%%%%%%%%%%%%%%%%%%%%%%%%%%%%%%%%%%%%%%%%%%%%%%
\begin{figure}[t]
\begin{center}
\epsfig{file=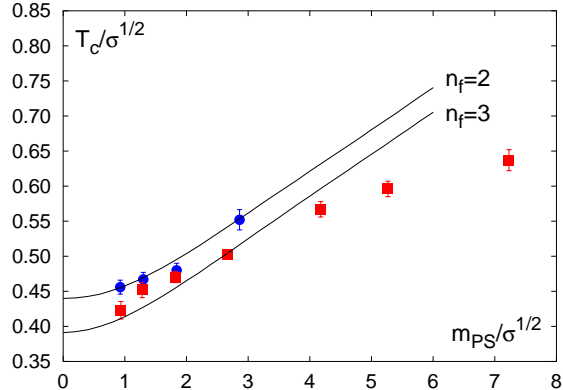,width=8.0cm}
\end{center}
\caption{Transition temperatures in 2 and 3 flavor QCD versus the
lightest pseudo-scalar meson mass. Both quantities are expressed
in units of the square root of the string tension. The solid lines
correspond to temperatures at which the energy density of a 2 and 3 
flavor resonance gas attains the value $\epsilon = 0.8$/GeV/fm$^3$.
The former consist of hadrons without strangeness content while the latter
also includes strange mesons and baryons.
\label{tc23}}
\end{figure}
%%%%%%%%%%%%%%%%%%%%%%%%%%%%%%%%%%%%%%%%%%%%%%%%%%%%%%%%%%%%%%%%%%%%%%%

The good agreement between lattice results on the QCD equation of 
state in the low temperature phase and the resonance gas model
raises the question whether we can also understand the dependence of 
the QCD transition temperature on the quark mass in terms of a resonance 
gas. The similarity of the ''critical'' energy densities found 
in the infinite quark mass limit and for (moderately) light quark masses
suggests that the transition to the quark gluon plasma phase occurs 
at approximately constant energy density, irrespective of the quark mass
values or the physical masses of the hadronic resonances. This assumption
is the basis for the comparison of transition temperatures in 
2 and 3 flavor QCD with lines of constant energy density calculated
in a resonance gas model\cite{res1}. Results are shown in Fig.~\ref{tc23}. 
Up to pseudo-scalar masses $m_{PS}\simeq (3-4)\sqrt{\sigma} \simeq
(1.3-1.7)$~GeV the agreement with the hadronic resonance gas is quite
reasonable. For larger pseudo-scalar masses the glueball sector does start to
play an increasingly important role as the heavy hadrons decouple from
the thermodynamics and the ''lighter'' glueballs yield the
largest contribution to the energy density. Aside from the inclusion
of glueball states in the resonance gas\cite{res1} one eventually also has 
to take into account that thermal effects may strongly influence the
glueball spectrum close to the transition temperature\cite{suganuma}.
Doing so the qualitative features of the quark mass dependence of $T_c$
can be modeled in the entire mass regime.

Finally, let us consider the lines of constant energy density in the 
$T$-$\mu_B$ plane. Separating the meson and baryon contribution to the 
energy density,
\begin{equation}
\epsilon (T,\mu_B) = \epsilon_M (T) + \epsilon_B (T) \cosh(\mu_B / T)
\quad,
\label{energy}
\end{equation}
where $\epsilon_B (T)$ is the baryonic contribution to the 
energy density at $\mu_B = 0$, one now can follow the strategy applied 
also in the lattice calculations and determine the lines of constant energy 
density from a leading order Taylor expansion. Expanding around the
transition point at vanishing chemical potential, $(T_c,\mu_B=0)$, yields,
\begin{equation}
\frac{T_c(\mu_B)}{T_c} = 1 - \frac{1}{2}
\frac{\epsilon_B(T_c)}{T_c \frac{\partial}{\partial T} \left( \epsilon_M (T)
+\epsilon_B(T) \right)_{T=T_c} }
\left( \frac{\mu_B}{T} \right)^2 \quad .
\label{constant}
\end{equation}
We note that the denominator on the right hand side is just the specific
heat in a resonance gas at $(T_c,\mu_B=0)$. This shows that one does expect
a quite weak dependence of the transition temperature on the chemical
potential, if the transition at $\mu_B = 0$ is close to 
a second order transition. Although this is in qualitative agreement
with the weak $\mu_B$-dependence of the transition line observed
in lattice calculations it seems that the resonance gas still leads
to a somewhat stronger variation with $\mu_B$. A more direct comparison in the 
limit of physical quark masses would be desirable here.

\section{Conclusions}

We have shown that the hadronic resonance gas model is able to 
describe quite a few quantitative and qualitative results obtained
in lattice calculations on the thermodynamics of the low temperature
hadronic phase of QCD as well as basic properties of the transition line
to the high temperature phase. This suggests that the copious production
of hadronic resonance indeed plays an important role in triggering the
transition to the quark gluon plasma phase of QCD. Unfortunately, a
direct comparison of lattice calculations with an non-interacting gas of 
resonances given directly in terms of the experimentally known hadron 
spectrum is at present not yet possible. 
If one wants  to compare the current lattice calculations, which still 
are being performed with too heavy quark masses, with the hadronic
resonance gas some phenomenological input on the quark mass dependence
of hadron resonances is required. However, it will soon be 
possible to further reduce or even eliminate the present ambiguities. 
With the availability 
of a new generation of Teraflops computers for lattice QCD it soon will
become possible to perform studies of QCD thermodynamics with an ''almost''
realistic spectrum of quark masses.

\vspace{0.3cm}
\section*{Acknowledgements}
This paper summarizes talks given at the conferences on ''Finite Density QCD''
and ''Confinement 2003'' held in July 2003 in Nara and Wako, Japan, respectively. 
I would like to thank 
the organizers, in particular A. Nakamura and H. Suganuma, for the hospitality 
extended to me during these very exciting and stimulating meetings. Attending 
these conferences has also been made possible through a travel grant of the German
Research foundation (DFG).

\vfill\eject

\end{document}